\documentclass[preprintnumbers, twocolumn,showkeys,superscriptaddress,showpacs,
nofootinbib,fleqn,prd,notitlepage]
{revtex4-1}

\usepackage{bbm,pifont}
\usepackage{amssymb}
\usepackage{amsmath}
\usepackage{physics}
\usepackage{amsfonts}
\usepackage{epsfig}
\usepackage{graphicx}
\usepackage{tabularx}
\usepackage{color}
\usepackage{hyperref}
\usepackage{slashed}

\usepackage[running]{lineno}

\begin{document}

\title{Probing quantum chaos with the entropy of decoherent histories}	

\author{Evgeny Polyakov}
\email[]{evgenii.poliakoff@gmail.com}
\affiliation{Russian Quantum Center, 30 Bolshoy Boulevard, building 1, Skolkovo Innovation Center territory, Moscow, 121205, Russia} 
\author{Nataliya Arefyeva}
\email[]{arefnat8@gmail.com}
\affiliation{Russian Quantum Center, 30 Bolshoy Boulevard, building 1, Skolkovo Innovation Center territory, Moscow, 121205, Russia}
\affiliation{Physical Department, Lomonosov Moscow State University, Vorobiovy Gory, Moscow 119991, Russia}

\begin{abstract}

Quantum chaos, a phenomenon that began to be studied in the last century, still does not have a rigorous understanding. By virtue of the correspondence principle, the properties of the system that lead to chaotic dynamics at the classical level must also be present in the underlying quantum system. In the classical case, the exponential divergence of nearby trajectories in time is described in terms of the Lyapunov exponent. However, in the quantum case, a similar description of chaos is, strictly speaking, impossible due to absence of trajectories. There are different approaches to remedy this situation, but the universal criterion of quantum chaos is absent. We propose the quantum chaos definition in the manner similar to the classical one using decoherent histories as a quantum analogue of trajectories. For this purpose, we consider the model of an open quantum kicked top interacting with the environment, which is a bosonic bath, and illustrate this idea. Here, the environment plays the role of a trajectory recording device. For the kicked top model at the classical level, depending on the kick strength, crossover occurs between the integrable and chaotic regimes. We show that for such a model, the production of entropy of decoherent histories is radically different in integrable and chaotic regimes. Thus, the entropy of an ensemble of quantum trajectories can be used as a signature of quantum chaos.
\end{abstract}

\maketitle

\section{Introduction}
Chaotic behavior plays a significant role in various fields of science (for example, it underlies classical thermodynamics \cite{Zaslavsky2001, Zaslavsky:81, Krylov:79} and hydrodynamics \cite{Gaspard:97}). In classical systems, chaos is characterized by the exponential sensitivity of the evolution of the system in time to initial conditions, but in quantum mechanics, it is not possible to characterize chaos in the same way, since the concept of phase space trajectories loses its meaning due to the Heisenberg uncertainty principle. There are different approaches to the definition of quantum chaos: through the statistics of energy levels \cite{Robnik:98,Bohigas:1983er,Haake2010,Lozej2020}; spectral form factors \cite{Haake2010}; Loschmidt echo \cite{LE:862490, Jacquod2009}; out-of-time ordered correlators (OTOC) \cite{Rozenbaum2017,Xu2022,Hashimoto2017}; the rate of increase of entropy \cite{Zurek:94,Schack:96}; in the context of quantum modeling through fidelity decay \cite{Emerson2002} and others. However, the true understanding of the nature of quantum chaos and the limits of using its various diagnostics, as well as the possible connection between them, is the subject of ongoing research, both theoretical and experimental. Currently, it is impossible to present a universal criterion for determining quantum chaos and to rigorously understand this phenomenon. The methods of diagnosing quantum chaos have their drawbacks. For example, level statistics are poorly defined for small systems, and there are specific examples for which it does not work \cite{Bogomolny:97}, OTOC does not work for billiard systems, and in this case, it is not possible to distinguish integrable behavior from chaotic \cite{Hashimoto2017}. Thus, the interest in finding universal criteria for quantum chaos for classically chaotic systems, as well as understanding the nature of the appearance of this phenomenon is motivated.

Interest in quantum chaos is caused by its wide application in explaining fundamental problems, such as: the thermalization mechanism in isolated systems, for which the eigenstates of quantum chaotic systems play a significant role \cite{DAlessio2016,Deutsch2018,Srednicki:94}; quantum information scrambling \cite{Xu2022}; in relation to open quantum systems, the influence of chaos on the processes of decoherence and dissipation \cite{Marchiori2011,Xavier2015,Wen-Lei2010,Mirkin2021,Blume-Kohout2003}, etc. Currently, there are various experimental realizations of chaotic behavior, such as spin chains \cite{Braun2020,gubin246} implemented with cold atoms \cite{chaudhury461,Neill2016}.

In this work, we rely on the idea of Berry \cite{Berry2001}, who stated the importance of the environment for the emergence of quantum chaos. Quantum decoherence that occurs in non-isolated systems inhibits the quantum suppression of chaos (due to the fact that quantum systems have discrete, quantized energy levels that control the evolution of dynamic quantities; therefore, this evolution cannot be truly chaotic). Due to the environment, it is possible to introduce the concept of quantum trajectories of a system as a record that is stored in certain degrees of freedom of the environment\cite{Brun2004}. 

\begin{figure*}[!htb]
\includegraphics[width=0.9\textwidth]{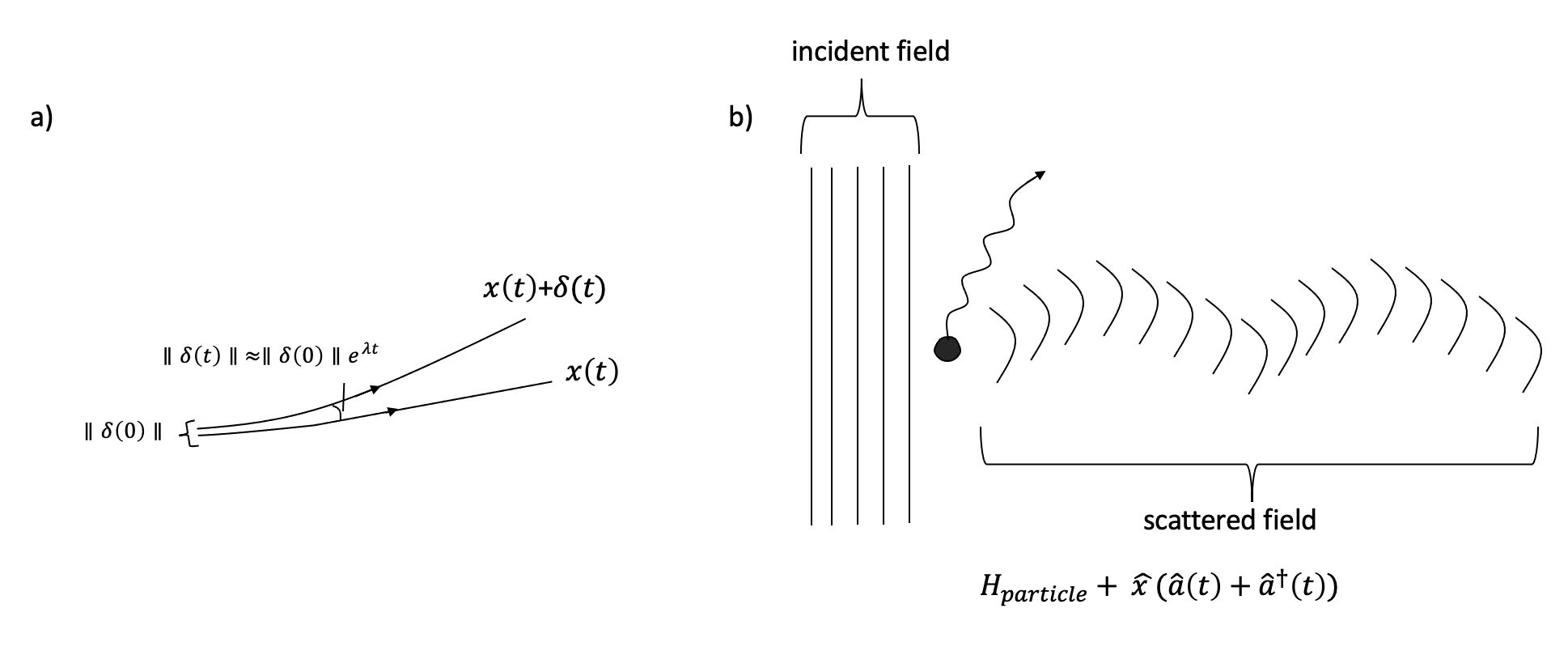}
\caption{a) Classical Lyapunov exponent through classical trajectories; b) The incident field is scattered by particles. A quantum analog of the trajectory is encoded in the scattered field, which can be modeled by coupling the target particle to the quantum environment through the environment’s operator $\hat{a}(t)$.\label{traject}}
\end{figure*}

We consider a model where a quantum environment is connected to an open quantum system (OQS), which in some degrees of freedom records how the system behaves as it evolves over time. This is similar in spirit to the decoherent histories approach, also known as consistent histories approach
\cite{Halliwell:95,Griffiths:93,Hohenberg2012}. Therefore, we call the recorded information about the OQS  decoherent history. The approach of decoherent histories in relation to quantum chaos can be found in the works \cite{Brun:95,Brun:96,Brun:93}, in which quantum dissipative chaotic systems are studied in Markov approximation, providing a connection with the classical limit. In this paper we propose a general treatment of quantum chaotic systems in the decoherent histories approach beyond Markov approximation. 

To correctly determine the decoherent histories, it is necessary to identify the degrees of freedom that carry information about how the OQS moved in the past. 
The formalism developed in this paper consists of several stages. First, the environmental degrees of freedom (later we can call them modes) that can carry information about the OQS are determined. There are infinitely many degrees of freedom in the environment, but only those degrees of freedom that have significantly interacted with the OQS can carry useful information. To achieve this, it is convenient to introduce the Lieb-Robinson light cone formalism \cite{Lieb:72}, which describes the propagation of perturbation. The effectively interacting degrees of freedom will be inside the light cone. Second, from these degrees of freedom, the irreversibly decoupled ones are determined, since the trajectory record should not change at future times and should not depend on future OQS evolution. Another words, they must carry away information about the OQS and stop interacting. Knowing these degrees of freedom, we can measure them one after another, and the sequence of the measurement results is a quantum trajectory (decoherent history).

The analogue of the trajectory appears due to the fact that the system interacts with the environment Fig.\ref{traject}. The formation of quantum trajectories corresponds to the emergence of decoherent histories in the environment \cite{Brun:97,Brun2004}.

The approach used in this work is based on the method  \cite{Polyakov2022}, which allows modeling of OQS dynamics beyond the limits of the Markov approximation applicability \cite{Diosi2012,Breuer2002}. In this work, this approach is adapted and the environment modes, which contain information about the OQS motion, are microscopically derived. Consequently, the concept of decoherent histories is constructed and the entropy of the ensemble of quantum trajectories \cite{Slomczynski:94,Brun:99,Brun2004} is calculated. It is reasonable to assume that the entropy of the ensemble of these quantum trajectories will be radically different in the integrable and chaotic regimes, as proved in this study. 

The paper is structured as follows. Sections \ref{kickedTop} and \ref{OQS} introduce the model in question. In Section \ref{ModelEnv} we explain our treatment of the decoherence history approach. Section \ref{EnvDegF} describes a method for deriving the environmental degrees of freedom, which contain information about the OQS motion. In Section \ref{Sec6} we construct quantum trajectories (decoherent histories) and calculate the entropy of an ensemble of such trajectories. In Section \ref{Sec7} we present our results. We conclude in Section \ref{conclusion}.

\section{The Considered chaotic system\label{kickedTop}}

\begin{figure*}[!htb]
\includegraphics[width=0.4\textwidth]{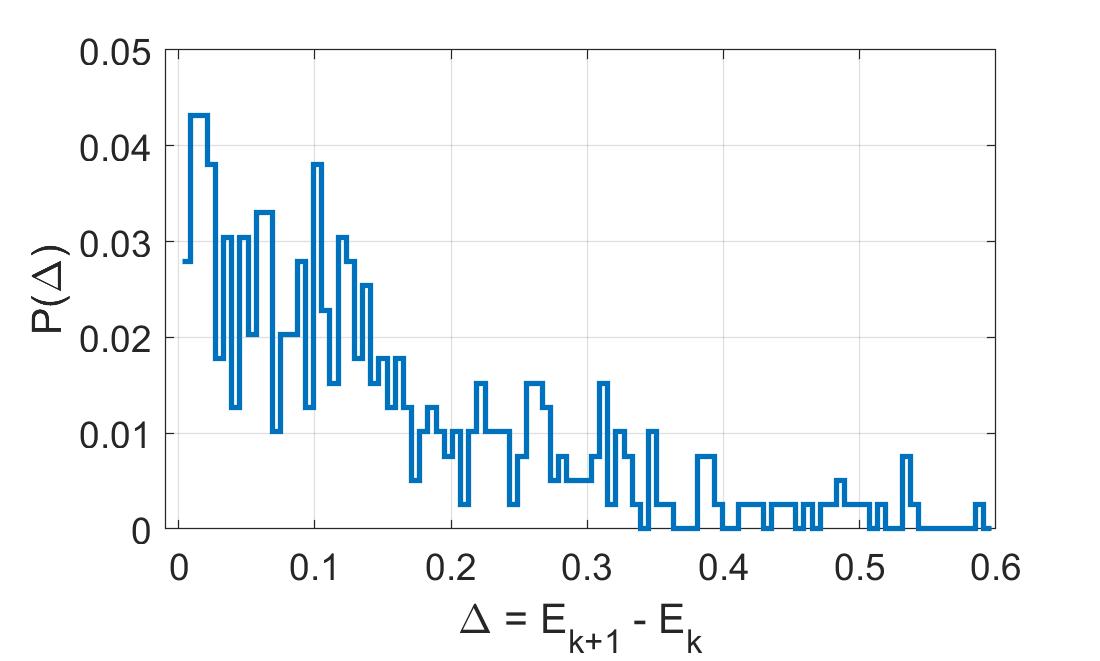}
\includegraphics[width=0.4\textwidth]{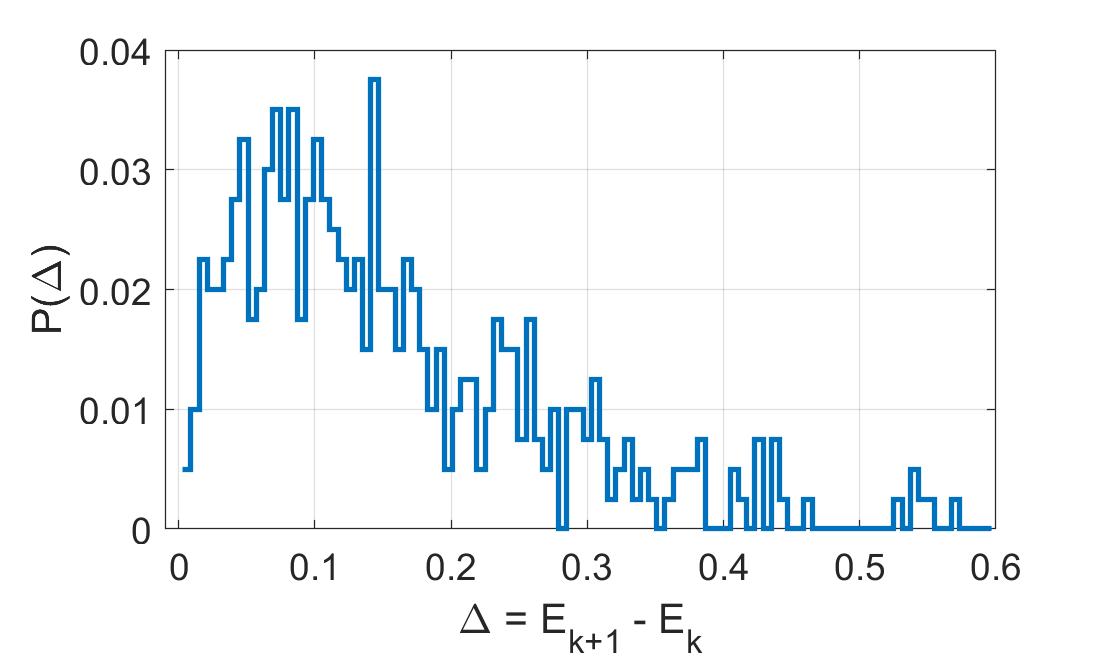}
\caption{Crossover between integrable (Poisson statistics) and chaotic (Wigner-Dyson statistics) motion; for the left statistics $K=2$, for right $K=3$. Image was plotted for $j=40$. In this work we use the following parameters for kicked top (eq. (\ref{kt})): $\beta=0.1$, $\tau=1$, $p=1.7$.
\label{crossover}}
\end{figure*}

We consider the model of a quantum kicked top \cite{Haake:87} as OQS, which at the classical level has chaotic behavior for certain values of kick strength $K$ (hereinafter, the natural system of units is used everywhere: $\hbar = 1$). This model has been well studied in the context of quantum chaos \cite{Haake2010,Wang:2023,Haake:87}:

\begin{equation}\label{kt}
\hat{H}_S = \dfrac{p}{\tau} \hat{J}_y + \dfrac{K}{2j} \left( \hat{J}_z - \beta\right)^2 \sum_{n = -\infty}^\infty \delta(t - n\tau)
\end{equation}

The system is characterized by the angular momentum $\vec{J} = \left(J_x,J_y,J_z\right)$ with the corresponding commutators: $[J_i,J_j] = i\epsilon_{ijk}J_k$ ($i,j,k$ run through $x,y,z$). The classical limit is reached by tending $j \rightarrow \infty$, $\hbar \rightarrow 0$ while preserving $\hbar j$. The first term is responsible for the precession around the $y$ axis with the angular frequency $\dfrac{p}{\tau}$, the second one is related to the periodic sequence of kicks at the time distance $\tau$.

By changing $K$, the motion of the system changes from integrable to chaotic. Fig.\ref{crossover} shows the level spacing distributions for different values of the kicked strength $K$.

The physical implementation of this model is provided by the system of interacting spins \cite{Waldner:85,chaudhury461}.

\section{Open chaotic quantum system\label{OQS}}
\begin{figure*}[!htb]
\centering
\includegraphics[width=1\textwidth]{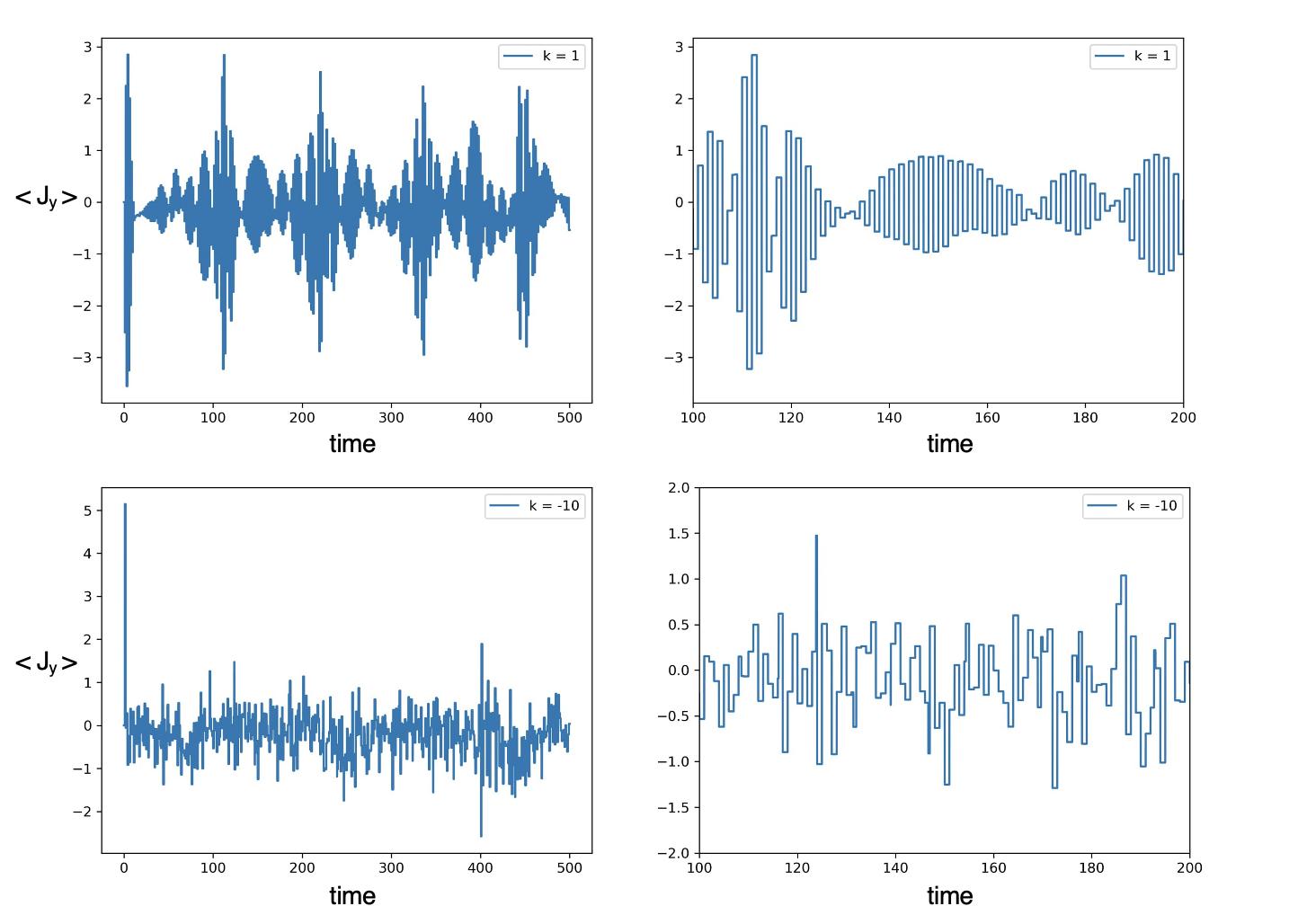}
\caption{Mean value of $J_y$ versus time along one trajectory. Top images are for regular motion $K=1$; lower images are for $K=-10$; the right plots enlarge the left ones. The initial condition $|\Psi(0)\rangle = |J_y = 0\rangle\otimes|0\rangle_E$, $|0\rangle_E$ is vacuum state of the environment. Images were obtained using the following parameters for the environment: $\epsilon_n=1$, $h_n=0.2$, $h=0.05$.
\label{mean_Jy_time}}
\end{figure*}

Our main idea is to introduce trajectories in the quantum case to obtain a method for diagnosing quantum chaos. To do this, it is necessary to connect the environment to the considered chaotic system (in this work the model of a quantum kicked top) (Section \ref{kickedTop}). The role of the environment is played by a bosonic bath.

The complete Hamiltonian of the system is as follows:
\begin{equation}
    \hat{H} = \hat{H}_S + \hat{H}_E + \hat{H}_{int} 
\end{equation}
where $\hat{H}_S$, $\hat{H}_E$, $\hat{H}_{int}$ are the free Hamiltonians of the OQS (\ref{kt}), the environment, and the Hamiltonian of the interaction between them, respectively, and are given by:
\begin{equation}
\hat{H}_E = \int\limits_{0}^\infty \omega\, \hat{a}^\dagger(\omega)\, \hat{a}(\omega)\, d\omega
\end{equation}

\begin{equation}
\hat{H}_{int} = \hat{J}_y\,(\hat{a}^\dagger + \hat{a}), \quad
\hat{a} = \int\limits_0^\infty c(\omega)\, \hat{a}(\omega) \,d\omega
\end{equation}
where $\hat{a}^\dagger(\omega)$, $\hat{a}(\omega)$ are the bosonic environment's creation and annihilation operators with $\left[\hat{a}(\omega), \hat{a}^\dagger(\tilde{\omega}) \right] = \delta (\omega - \tilde{\omega} )$, and $c(\omega)$ is the coupling. Such an interaction indicates that the environment records the trajectory of the projection of the y-component of the angular momentum of the kicked top

Fig.\ref{mean_Jy_time} shows the behavior of a quantum kicked top in the case of integrable and chaotic motion. In the following sections, we describe how these results were obtained.

In the interaction picture with respect to the free bosonic environment:
\begin{equation}
\hat{H}(t) = \hat{H}_S(t) + \hat{J}_y\,(\hat{a}^\dagger(t) + \hat{a}(t))
\end{equation}
\begin{equation}
\hat{a}(t) = \int\limits_0^\infty c(\omega)\, \hat{a}(\omega) e^{-i\omega t} d\omega
\end{equation}

In our work, it is convenient to represent the environment in the equivalent chain representation \cite{Chin:2010}. This representation is necessary in order to introduce the concept of the Lieb–Robinson light cone \cite{Lieb:72}. For a sufficiently wide class of spectral densities, there exists an unitary operator $U$ that takes the system into a chain representation \cite{Chin:2010}. Using the unitary operator, the environment is represented as a chain in which only neighboring modes interact:
\begin{equation}
    a_n^\dagger = \int\limits_0^\infty U_n(\omega)\, \hat{a}^\dagger(\omega)\, d\omega 
\end{equation}

\begin{multline}\label{chain}
\hat{H}(t) = \hat{H}_S(t) + \hat{J}_y \, h \, ( \hat{a_0}^\dagger  + \hat{a_0} ) + \\
+\sum_{n=0}^\infty \left( \epsilon_n \hat{a}_n^\dagger \hat{a}_n + h_n \hat{a}_{n+1}^\dagger \hat{a}_n + h_n \hat{a}^\dagger_n \hat{a}_{n+1}\right)
\end{multline}
with commutator $[\hat{a}_i,\hat{a}_j^\dagger]=\delta_{ij}$. Knowing the spectral density, the coefficients $\epsilon_n$, $h_n$ and $h$ can be calculated by recurrent formulas using orthogonal polynomials \cite{Chin:2010}. 

In the interaction picture with respect to the free bosonic environment in chain representation, we obtain the following:
\begin{equation}
\hat{H}(t) = \hat{H}_S(t) + \hat{J}_y \,h \,( \hat{a_0}^\dagger(t)  + \hat{a_0}(t) )
\end{equation}

\section{Quantum environment as the recorder of OQS trajectories\label{ModelEnv}}

In this study, the main idea is to consider the environment as a recording device that records information about the movement of the OQS in some degrees of freedom. Thus, the environment contains a sequence of projection operators corresponding to the records (facts) of the OQS motion. The definition of the trajectories we introduce is related to the approach of decoherent histories \cite{Halliwell:95,Griffiths:84,Hohenberg2012}. 

Consistent histories, also known as decoherent histories (DH) formalism, were introduced by Griffiths, Omnes, Gell-Mann, and Hartle \cite{Griffiths:84, Gell-Mann:93}. This formalism is an interpretation of quantum mechanics that allows one to resolve/tame the main quantum paradoxes. The DH approach is based on the assumption of the probabilistic nature of quantum time dependence \cite{Hohenberg2012}. The DH concept also aims to explain how classical reality emerges from quantum mechanics \cite{Gell-Mann:93}.

A “history” is a set of events or prepositions, represented by projection operators $\hat{P}^1_{\alpha_1},...,\hat{P}^n_{\alpha_n}$ at a succession of times $t_1,...,t_n$ time-ordered with unitary evolution between each projection. At each time moment $t_i$, there are different alternatives $\alpha_i = 1,...,m_i$ that correspond to a set of projection operators $\{\hat{P}^i_{\alpha_i}\}$, $\alpha_i$ numbers particular alternatives and $i$ numbers time moments. Such sets are exhaustive and exclusive at each time moment $t_i$: (i) the sum of all alternatives of the set is unity $\sum_{\alpha_i} \hat{P}^i_{\alpha_i} = \mathbbm{1}$; (ii) two distinct alternatives are mutually orthogonal $\hat{P}^i_{\alpha_i}\hat{P}^i_{\beta_i} = \delta_{\alpha_i \beta_i}\hat{P}^i_{\alpha_i}$. Thus the history is represented as time-ordered product of the projection operators \cite{Halliwell:95,Gell-Mann:93,Halliwell:93}:

\begin{equation}\label{proj}
\hat{C}_{\alpha_1,...,\alpha_n} = \hat{P}^n_{\alpha_n}(t_n)\hat{P}^{n-1}_{\alpha_{n-1}}(t_{n-1})...\hat{P}^1_{\alpha_1}(t_1)
\end{equation}
where
\begin{equation}
\hat{P}^i_{\alpha_i}(t_i) = e^{\frac{i}{\hbar}\hat{H}_E(t_i-t_{i-1})}\hat{P}^i_{\alpha_i} e^{-\frac{i}{\hbar}\hat{H}_E(t_i-t_{i-1})}
\end{equation}

Here the projections correspond the records inside degrees of freedom of the bath. We consider a bipartite system comprising an OQS and a bosonic bath, $|\Psi\rangle$ is wave function in joint Hilbert space of OQS and bath, the density matrix is $\rho = |\Psi\rangle\langle\Psi|$. 

The probability of a history, a sequence of alternatives, can be given as \cite{Halliwell:95}:
\begin{equation}
p(\alpha_1,...,\alpha_n) = \Tr(\hat{C}_{\alpha_1,...,\alpha_n} \, \rho \,\hat{C}_{\alpha_1,...,\alpha_n}^\dagger)
\end{equation}
However, in general, it does not obey all the probability sum rules, e.g. $p(\alpha_2) = \sum_{\alpha_1} p(\alpha_1,\alpha_2)$. For satisfying them, the following condition is necessary and sufficient \cite{Griffiths:84,Gell-Mann:93}: 
\begin{equation}\label{consistent}
\Re \Tr(\hat{C}_{\alpha_1,...,\alpha_n}\rho\,\hat{C}_{\beta_1,...,\beta_n}^\dagger) = 0 \, ,\quad \text{for $\alpha\neq \beta$}
\end{equation}

In this paper we use a stronger consistency condition, which means that all different pairs of histories $\alpha$, $\beta$ do not interfere (medium decoherence condition \cite{Gell-Mann:93}):
\begin{equation}\label{consistent2} \Tr(\hat{C}_{\alpha_1,...,\alpha_n}\rho\,\hat{C}_{\beta_1,...,\beta_n}^\dagger) = 0 \, ,\quad \text{for $\alpha\neq \beta$}
\end{equation}

In practice, it is generally difficult to achieve consistency condition \cite{Arrasmith:2019} and find exactly decoherent sets. However, the consistency condition can be approximated arbitrarily well with respect to a given level of significance \cite{Halliwell:95}. Thus, we arrive at the condition of approximate consistency:
\begin{equation}\label{weak_consistent}
\Tr(\hat{C}_{\alpha_1,...,\alpha_n}\rho\,\hat{C}_{\beta_1,...,\beta_n}^\dagger) \approx 0 \, ,\quad \text{for $\alpha\neq \beta$}
\end{equation}

In the DH approach, the question arises of how to build these projections $\hat{P}^i_{\alpha_i}$ and what observables and time moments to consider. There is some arbitrariness in the choice of this \cite{Slomczynski:94}. Moreover, it is difficult to construct them. Recently, it was proposed to search for them on a quantum computer \cite{Arrasmith:2019}. Our approach naturally resolves this problem. On the one hand, we have a physical model in which the scattered field carries away information about the OQS motion, and on the other hand, we propose a formal consideration of how these projections may be found. The projections must match the degrees of freedom of the environment and naturally arise from the properties of the environment. In the next section, we derive these degrees of freedom.

\section{Environment degrees of freedom which carry the information about the trajectory\label{EnvDegF}}
In this section, we describe our procedure for deriving the environmental degrees of freedom, carrying useful information about the OQS trajectory.

\subsection{Statistically significant interacting modes}

The quantum environment is treated as a recording device. Its records can be measured, and decoherent histories can be obtained. Decoherent histories can only be contained in environmental modes within a light cone. Naively to find these modes it is necessary to solve a many-body problem. However, solving a many-body problem is difficult, and from a practical point of view, it is a useless approach. In our work, we first propose to approximate the light cone a priori and then solve the many-body problem inside it. Below is an algorithm for estimating the light cone a priori. 

The light cone allows one to determine which degrees of freedom are significant and which are not. The region outside the light cone consists of degrees of freedom that will only be significantly excited in the future or will never be excited at all. In particular, for each chain site in eq.(\ref{chain}), there is a point in time after which it becomes statistically significant for the evolution of the system.

In order to a priori estimate which modes the OQS excites, it is necessary to introduce a measure that determines the influence of the OQS on the considered mode. For this purpose, it is convenient to use the commutator $[\hat{a}_0(t), \hat{a}_j^\dagger]$, which will show whether the operator $\hat{a}_0(t)$ affects the mode corresponding to $\hat{a}_j$. If this mode is currently interacting with the OQS, then $\hat{a}_j$ and $\hat{a}_0(t)$ do not commute. The operator $\hat{a}_0(t)$, which is the degree of freedom with which the OQS interacts at time $t$ in the interaction picture, can be expressed in terms of the original chain operators as follows:
\begin{equation}
\hat{a}_0(t) = \sum_{k=0}^\infty \phi_k(t) \hat{a}_k
\end{equation}
where $\phi_k(t)$ is one-particle wave function that satisfies the following first-quantized Schrödinger equation with the initial condition corresponding to the interaction quench at time $t=0$:
\begin{equation}
\begin{cases}
\partial_t \,\phi_k(t) = - i \epsilon_k \phi_k(t) - i h_k \phi_{k+1}(t) - i h_{k-1} \phi_{k-1}(t)
\\
\phi_k(0) = \delta_{k0}
\end{cases}
\end{equation} 
where $h_{-1} \equiv 0$.

The Hamiltonian responsible for the evolution of the one-particle wave function evolution is as follows:
\begin{equation}
H_1 =
\left(
\begin{array}{cccccc}
\epsilon_0 & h_0 & 0 & \ldots & \ldots & 0  \\
h_0 & \epsilon_1 & h_1 &  0 &  \ldots &  \ldots \\
0 & h_1 & \epsilon_1 & h_2 & \ldots & \ldots  \\
\vdots & \vdots & \vdots & \vdots & \ddots & \vdots\\
0 & \ldots & \ldots & 0 & h_{m(t)-1}& \epsilon_{m(t)}
\end{array}
\right)
\end{equation}

\begin{figure}[!h]
\centering
\includegraphics[width=0.44\textwidth]{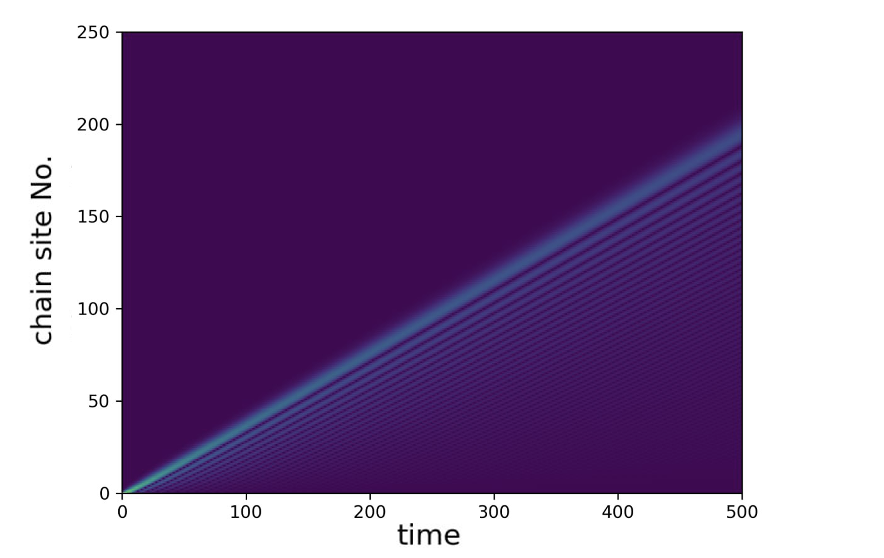}
\caption{Wave function $\phi_k(t)$ propagating the interaction operator $a_0(t)$ over time. The color matches $|\phi_k(t)|$. It can be seen that the perturbation propagates along the light cone.}
\label{oper_spread}
\end{figure}

Here $m(t)$ is the number of environmental degrees of freedom that have been excited due to co-evolution with the OQS over time $t$. The perturbation propagates along the Lieb-Robinson light cone \cite{Lieb:72} from the zero site $a_0$, with which the OQS is connected. Fig.\ref{oper_spread} shows the spread of the operator $\hat{a}_0(t)$ over sites of chain. 

For the simplest case of a linear environment, the commutator is:
\[
[\hat{a}_0(t), \hat{a}_j^\dagger] = c_j(t) \mathbbm{1}
\]
where $\mathbbm{1}$ is identity operator in bath Hilbert space.

Thus, the measure characterizing the instant interaction intensity with the OQS at time $t$ can be expressed as follows:
\begin{equation}\label{OTOC_DEF}
|c_j(t)|^2 = \langle0|[\hat{a}_0(t), \hat{a}_j^\dagger] [\hat{a}_0(t), \hat{a}_j^\dagger]^\dagger|0\rangle \equiv C_j(t)
\end{equation}
If $C_j(t)=0$, then mode $\phi_j$ does not interact. If $C_j(t)>0$, then mode $\phi_j$ is currently interact with OQS. This function takes the form of OTOC \cite{Xu2022}. The condition $C_j(t)>0$ indicates that the mode is coupled with the OQS at a given time. If the instant interaction intensity $C_j(t)$ is negligible, the excitation of this mode due to the OQS is also negligible.

The light cone is determined by the average intensity of the mode interaction over the time interval from $0$ to $t$ rather than by the instantaneous intensity of the mode interaction. During time $t$, only those modes enter the light cone that interact significantly on average over the entire interval. Therefore, it is necessary to consider only statistically significant interactions during the chosen time interval and eliminate sudden short-term excitations of environmental modes, which will make a negligible contribution.  The OTOC (\ref{OTOC_DEF}) averaged over time is as follows: 
\begin{equation}\label{OTOC_0t}
\langle C^+_j(t) \rangle \equiv \int \limits_0^t C_j(\tau) d \tau
\end{equation}

Since the boundary of the light cone is fuzzy, it always has exponentially decaying tails outside its front, so it is necessary to introduce a significance threshold where we cut off. Thus, the condition for the mode to be inside the light cone is that the average intensity of the mode interaction is above a certain threshold of significance $a_{cut}$: 
\begin{equation}\label{OTOC_0t_criterion}
\langle C^+_j(t)\rangle - a_{cut} > 0
\end{equation}
and we consider the modes that are effectively coupled and interact with OQS, influencing their joint evolution.

\subsection{Records may be nonlocal}

When constructing decoherent histories, we need to define the grid of times $t_i$ at which projections $\hat{P}_{\alpha_i}^i$ are appearing. In literature, one usually introduces time coarse graining ad-hoc by hands \cite{Brun:95, Brun:96}. In our case of non-Markovian quantum dynamics, the grid of times emerges naturally from the environmental spectral density of states. 

The light cone defined in a chain basis has a drawback, namely that the environmental modes are not statistically independent. 
Considering the environment as a transmission line, there is Kotelnikov sampling theorem \cite{Kotelnikov:93}, which states that statistically independent wave packets can be emitted into the line at a rate proportional to the bandwidth of this line. By analogy in the context of decoherent histories, truly statistically independent degrees of freedom will appear in a basis where the speed of propagation of the light cone is minimal, and the intervals between the times of appearance of modes are proportional to the width of the spectral density of the environment as the bandwidth of the environment as a recording device \cite{Polyakov2022}. The projections $\hat{P}_{\alpha_i}^i$ which carry independent bits of history, should occur in these times.

Therefore, it is necessary to generalize the concept of the light cone to an arbitrary frame. Instead of the chain operators $\hat{a}_i$ and $\hat{a}_i^\dagger$, we consider a unitary transformed set of them: $\hat{\kappa}_j^\dagger = \sum_{k=0}^\infty U_{jk} \hat{a}_k^\dagger$ for arbitrary unitary matrix $U$. 
Given a sequence of $\hat{\kappa_j}$, we can, by analogy with (\ref{OTOC_0t_criterion}) define the criterion when the mode corresponding $\hat{\kappa}_j$ enters the interaction with OQS for the first time. The average statistical significance of state $|\kappa_j\rangle = \sum_{k=0}^\infty U_{jk}|k\rangle$, where $|k\rangle$ is quantum localized in $k$ chain site, is as follows:
\begin{eqnarray}\label{eq20}
\langle C_{\kappa_j}^+(t)\rangle &=& \int \limits_0^t \langle 0|[\hat{a}_0(\tau), \sum_k U_{jk}\hat{a}_k^\dagger] [\hat{a}_0(\tau), \sum_l U_{jl}\hat{a}_l^\dagger]^\dagger|0\rangle d \tau \nonumber \\ 
&=& \langle\kappa_j| \int \limits_0^t d \tau |\phi(\tau)\rangle \langle\phi(\tau)|\kappa_j\rangle  = \langle\kappa_j| \rho_+(t)|\kappa_j\rangle 
\end{eqnarray}
with
\begin{equation}
 \rho_+(t) = \int \limits_0^t d \tau |\phi(\tau)\rangle \langle\phi(\tau)|
\end{equation}

\begin{figure}[!h]
\centering
\includegraphics[width=0.44\textwidth]{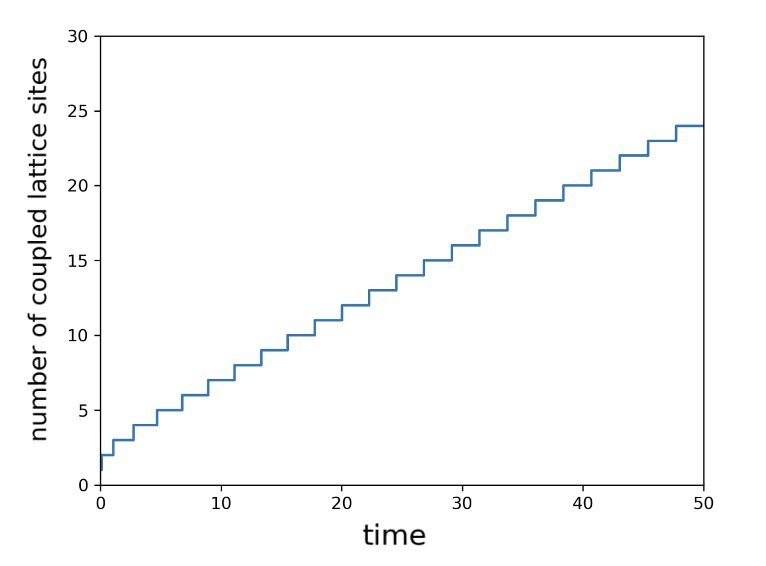}
\caption{The chain sites coupled to the OQS, depending on time, form a forward light cone. Coupled modes are defined according to eq.(\ref{LC}).}
\label{forwardLC}
\end{figure}

We introduce a metric that determines whether the contribution of the $|\kappa_j\rangle$ state is significant or not:
\begin{equation}\label{LC}
  g_+(\kappa_j,t) = \langle\kappa_j| \rho_+(t)|\kappa_j\rangle - a_{cut}  
\end{equation}
If $g_+(\kappa_j,t)<0$, the contribution of this mode can be neglected with threshold $a_{cut}$.
Modes lying inside the light cone, i.e. satisfying condition $g_+(\kappa_j,t) > 0$, contain information about the OQS (the kicked top). Fig.\ref{forwardLC} shows the modes (chain sites) coupled to the OQS over time, as determined according to eq.(\ref{LC}).

Since the Lieb-Robinson metric (\ref{LC}) is defined for an arbitrary mode, we can pose a variational problem and find $U$ where the light cone propagates with a minimum speed, which we call the minimal light cone $\hat{\kappa}_{l}^{\dagger}$$^{in}$$ = \sum_k U_{lk} \hat{a}^\dagger_k $. Moments when modes enter the light cone are delayed as much as possible. Then, by analogy with Kotelnikov theorem, modes that managed to enter the minimum light cone carry useful information, and the rest do not have time to interact and do not carry a decoherent history; they do not need to be considered. Further, unless otherwise stated, we will work within the frame of the minimal light cone. Information about the OQS is recorded in nonlocal environmental degrees of freedom.

A detailed algorithm for obtaining the minimal light cone is derived in the work \cite{Polyakov2022}.
We denote the modes coupled to the OQS for the time interval $[0,T]$ as $\kappa_{1}^{in},...,\kappa_{m_{in}(T)}^{in}$ and the discrete times of their appearance as $t_1^{in},...,t_{m_{in}(T)}^{in}$.

The total joint state of the quantum kicked top and bosonic bath $|\Psi(t)\rangle$ effectively evolves with the Hamiltonian:
\begin{multline}
\hat{H}_{eff}(t) = \hat{H}_S(t) \,\,+ \\
+\sum_{l=1}^{m_{in}(t)} \left\{ \hat{J}_y \langle \phi(t)|\kappa_l^{in} \rangle \hat{\kappa}_l^{in} + \hat{J}_y \langle \kappa_l^{in}|\phi(t) \rangle \hat{\kappa}_l^{in\dagger}\right\}
\end{multline}
Entanglement between degrees of freedom is neglected when their statistical significance is below a certain threshold. 

\subsection{Irreversibly decoupled modes -- stable records}
Records carrying information about the OQS must be stable facts; therefore, it is necessary to consider modes that are irreversibly decoupled from the OQS.

Two different cases of outgoing (decoupled) modes are possible: (a) modes that have never interacted with the OQS and (b) modes that interacted with the OQS and were irreversibly decoupled from it. The first situation does not contain any information about the OQS, and we discard these modes from consideration. However, it is necessary to track the evolution of (b). The mode decoupled from the OQS at time $t_l^{out}$ must be a linear combination of $\kappa_1^{in},\, \kappa_2^{in},\, ...,\, \kappa_{m_{in}(t_l^{out})}^{in}$. That is they must be in the subspace of modes coupled to the OQS in the time interval $[0,t_l^{out}]$. It is these modes will store information about the trajectory of the OQS.

Similarly to eq.(\ref{eq20}), for outgoing modes, the measure of statistical significance at time $t$, which determines the decoupling of the mode from the OQS, is
\begin{equation}\label{OTOC_IDM}
\langle C_{\kappa_j}^-(t) \rangle = \langle\kappa_j| \int \limits_t^T d \tau |\phi(\tau)\rangle \langle\phi(\tau)|\kappa_j\rangle  = \langle\kappa_j| \rho_-(t)|\kappa_j\rangle 
\end{equation}
with
\begin{equation}
\rho_-(t) = \int \limits_t^T d \tau|\phi(\tau)\rangle \langle\phi(\tau)|
\end{equation}
A mode can be considered irreversibly decoupled if the OTOC averaged over future times is negligible.

The condition of lack of statistical significance for them is as follows:
\begin{equation}
g_-(\kappa^{in},t) = \langle\kappa^{in} | \rho_-(t)|\kappa^{in}\rangle - a_{cut} < 0 
\end{equation}

These modes can be found similarly to the coupled modes in the minimal light cone by some unitary rotation of the basis of the coupled modes $\kappa_{1}^{in},...,\kappa_{m_{in}(T)}^{in}$. We denote the irreversibly decoupled modes for the time interval $[0,T]$ as $\kappa_{1}^{out},...,\kappa_{m_{out}(T)}^{out}$ and the discrete times of their decoupling as $t_1^{out},...,t_{m_{out}(T)}^{out}$. Information about the OQS is contained in irreversibly decoupled modes that have previously interacted with it. For more details see work \cite{Polyakov2022}.

\subsection{Relevant modes} 

By the time $t_k^{out}$, when the k-st $\kappa_k^{out}$ mode is decoupled, there are modes that remain coupled with the OQS. We call these modes relevant modes because they are statistically significant for future evolution. For the time moment $t_k^{out}$ they are consisted of modes that once became coupled to the OQS before $t_k^{out}$ ($\kappa_1^{in},...,\kappa_{m_{in}(t_k^{out})}^{in}$) except for modes that managed to irreversibly decouple from the OQS by this time ($\kappa_{1}^{out},...,\kappa_{k-1}^{out}$). Thus, there are $m_{in}(t_k^{out})$ coupled modes and $k-1$ irreversibly decoupled modes, and their difference is the number of relevant modes $r(t_k^{out})$:
\begin{equation}
    r(t_k^{out}) = m_{in}(t_k^{out}) - k + 1
\end{equation}

The total system state $|\Psi(t)\rangle$ evolves over the time interval $[t_k,t_k^{out}]$, where $t_k$ is the time of the previous mode coupling/decoupling event with the relevant modes $\kappa_1^{rel},...,\kappa_{r(t_k^{out})}^{rel}$ \cite{Polyakov2022}.

\begin{figure}[!h]
\centering
\includegraphics[width=0.4\textwidth]{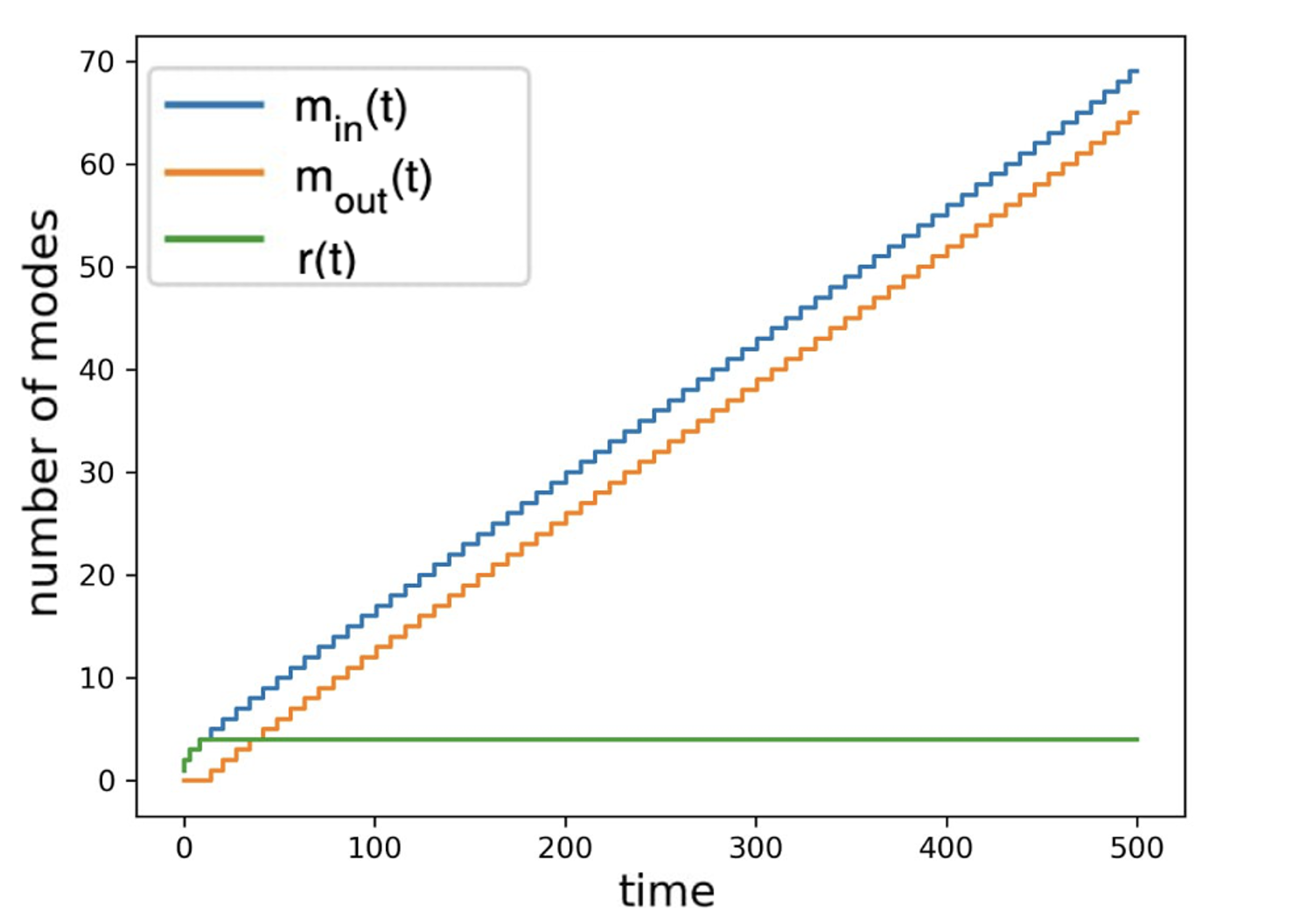}
\caption{The number of modes in the system over time, $m_{in}(t)$, $m_{out}(t)$, $r(t)$ --- coupled, irreversibly decoupled, and relevant modes, respectively. The entire time interval $T=500$.
\label{numberMode}
}
\end{figure}

Fig.\ref{numberMode} shows the number of coupled modes, irreversibly decoupled modes, and relevant modes over time. It can be seen that the number of relevant modes saturates and practically does not change during the evolution of the system.

There are no decoherent histories in relevant modes. After interaction quench, the OQS is renormalized (by analogy with the electron in high energy physics) and consists of a bare OQS and relevant modes with which it will interact significantly in the future. 

\subsection{Relation to DH approach}
The problem with the DH approach is related to the difficulty in achieving the consistency condition (\ref{consistent2}). Our approach suggests an effective solution to this problem. 

The records are contained in irreversibly decoupled modes; therefore, we find the projections $\hat{P}_{\alpha_k}^k$ in the subspace of these modes. The corresponding terms from the Hamiltonian are eliminated. The projections commute with the Hamiltonians; they become an integral of motion after time of irreversible decoupling mode and, thus, satisfy the sum rule, so the histories decohere.

\section{Simulating decoherent histories\label{Sec6}}
Knowing the degrees of freedom in which the environment records information about the trajectory of the kicked top, they can be measured. The measurement statistics will give an ensemble of quantum trajectories --- decoherent histories.

Before $t=t_k^{out}$ the mode $\kappa_k^{out}$ was coupled to the OQS. It was in an entangled state with the OQS due to Schmidt decomposition:
\begin{multline}\label{shmidt}
|\Psi(t_k^{out})\rangle =\sum_q c_q(k) \times\\
\times|\Psi_{coll}^{(q)}(t_k^{out})\rangle_{rel} \otimes |\Psi_{J}^{(q)}(t_k^{out})\rangle_{\kappa_k^{out}}
\end{multline}
where index $rel$ indicates belonging to a joint Hilbert space of OQS and relevant modes, and $\kappa_k^{out}$ refers to the newly formed irreversibly decoupled mode, $q$ enumerates the basis elements for such a mode.

Since this $\kappa_k^{out}$ mode is irreversibly decoupled, the amplitudes $c_q(k)$ do not depend on time; they are invariants. A sequence of motion invariants arises; they cease to effectively depend on time by the threshold of significance. Thus, form (\ref{shmidt}) is invariant at all future times, and an invariant entanglement structure arises for future evolution. This has also been confirmed numerically. By performing the Schmidt decomposition recursively at the moments of modes decoupling, the emerging invariant structure of entanglement is as follows:
\begin{multline}\label{EM_INV_STR}
|\Psi(t)\rangle =\sum_{q_1,...,q_k} c_{q_1}(1) c_{q_2}(2|q_1) ...c_{q_k}(k|q_1,...,q_{k-1}) \times \\
\times|\Psi_{coll}^{(q_1...q_k)}(t)\rangle_{rel} \otimes |\Psi_{J}^{(q_1)}(t_1^{out})\rangle_{\kappa_1^{out}} \otimes...\\
...\otimes |\Psi_{J}^{(q_k)}(t_k^{out})\rangle_{\kappa_{m_{out}(t)}^{out}} 
\end{multline}
It carries an ensemble of decoherent histories. 

According to the von Neumann measurement model \cite{Luder:2006}, one can collapse the wave function (\ref{shmidt}) and interpret the equation as the $k$-th quantum jump at time $t=t_k^{out}$: $|\Psi(t_k^{out})\rangle \rightarrow |\Psi^{(q)}_{coll}(t_k^{out})\rangle $
with probability $|c_q(k|q_1,...,q_{k-1})|^2$. Such quantum jumps are irreversible over time. 

By the time $t$, $m_{out}(t)$ modes have been irreversibly decoupled (\ref{EM_INV_STR}). Each mode decoupling is accompanied by a quantum jump, which is obtained from the measurement procedure recurrently applied:
\begin{eqnarray}
&&|\Psi(t_1^{out})\rangle \rightarrow |\Psi^{(q_1)}_{coll}(t_1^{out})\rangle_{rel} \nonumber\\
&&|\Psi^{(q_1)}_{coll}(t_2^{out})\rangle_{rel}  \rightarrow |\Psi^{(q_1q_2)}_{coll}(t_2^{out})\rangle_{rel} \\
&&|\Psi^{(q_1q_2)}_{coll}(t_3^{out})\rangle_{rel}  \rightarrow |\Psi^{(q_1q_2q_3)}_{coll}(t_3^{out})\rangle_{rel}\nonumber \\
&&...\nonumber
\end{eqnarray}

Therefore, $m_{out}(t)$ quantum jumps occur before time $t$. They are characterized by the history of choices $h = (q_1,q_2,…,q_k) = \{q_k\}_{k:\, t_k^{out}\leq t}$, appearing with probabilities:
\begin{equation}\label{dec_hist}
    P\,(q_1,q_2,…,q_k) \,\,\,= \prod\limits_{k:\, t_k^{out}\leq t} |c_{q_k}(k|q_1,...,q_{k-1})|^2 
\end{equation}

This is the proposed definition of decoherent histories. An important feature of our approach is that the average of all decoherent histories observables $h$ up to time $t$ corresponds to the full many-particle quantum dynamics of the OQS in terms of the significance threshold \cite{Polyakov2022}.

Thus, in the environment, projection operators (eq.(\ref{proj})) in the DH approach (Sec.\ref{ModelEnv}) naturally appear as follows:
\begin{equation}
\hat{P}_{\alpha_k}^k = \mathbbm{1}_{rel}\,\otimes\,|\Psi_{J}(t_k^{out})\rangle_{\kappa_k^{out}}\langle\Psi_{J}(t_k^{out})|_{\kappa_k^{out}}
\end{equation}

\subsection{The entropy of decoherent histories ensemble}
The statistical ensemble of quantum jump histories is encoded in an emerging invariant entanglement structure (\ref{EM_INV_STR}) that does not change in future. 

In summary, a measuring device is required to observe the trajectory. By adding the environment, which is considered a recording device, information about the trajectory is recorded in the stream of irreversibly decoupled degrees of freedom.

We now introduce the definition of the entropy of an ensemble of decoherent histories (\ref{dec_hist}):
\begin{equation}\label{entropy}
S = - \sum_{h = (q_1,...q_N)}P(q_1,...q_N) \ln P(q_1,...q_N)
\end{equation}

\section{The entropy of decoherent histories as a marker of quantum chaos\label{Sec7}}
We proposed that the entropy of a decoherent histories ensemble (\ref{entropy}) may be a criterion for quantum chaos. In this section, we present our main results. 

The entropy was calculated by considering the simplified assumption of the presence of ergodicity for quantum trajectories. In the sense that averaging over all trajectories is equivalent to averaging within one sufficiently long trajectory over all choices. Averaging over one trajectory was used in this study.

\begin{figure}[!h]
\centering
\includegraphics[width=0.43\textwidth]{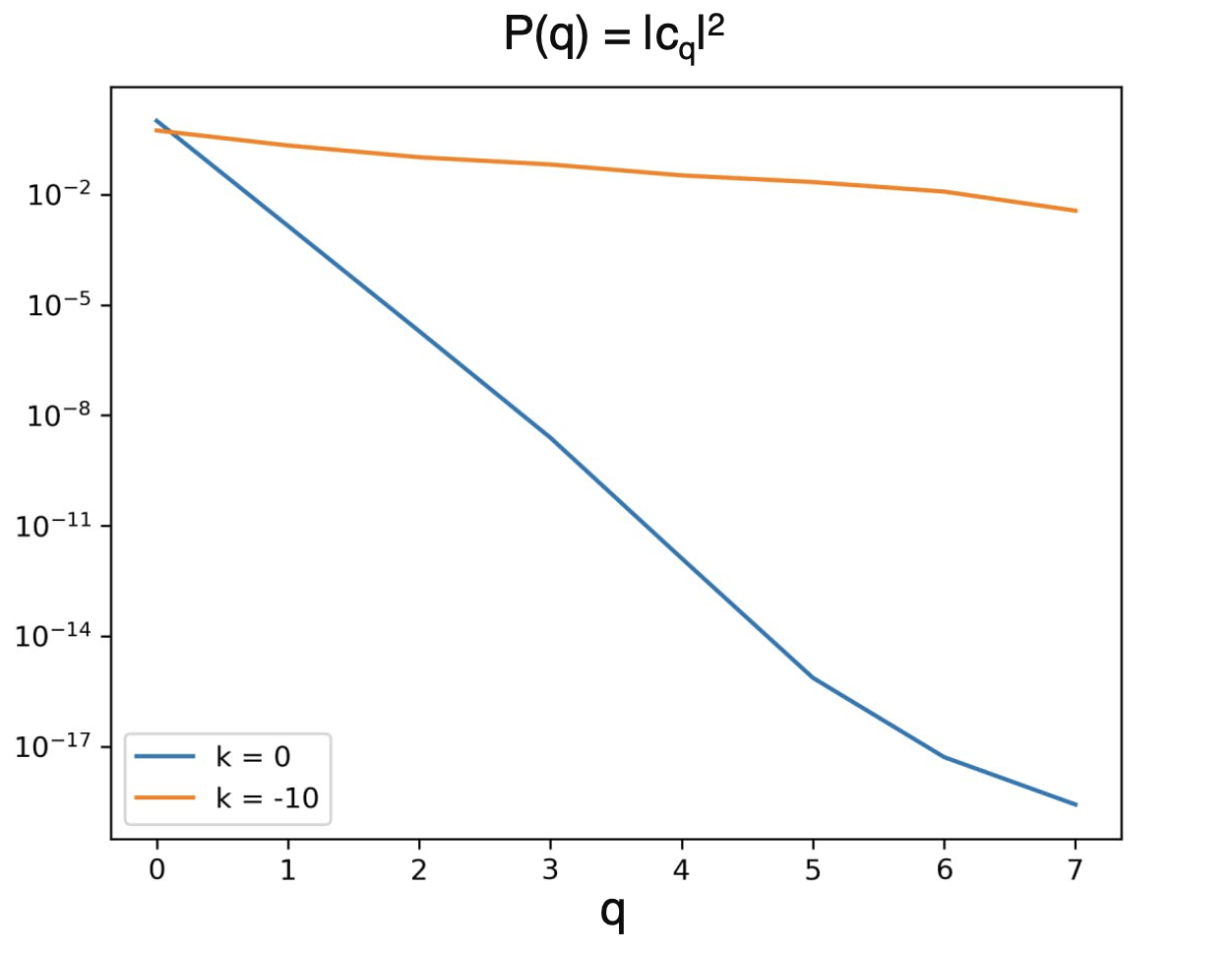}
\caption{Quantum jump probability distribution $|c_q|^2$ in two cases for kick strength $K=0$ (blue curve) and $K=-10$ (orange curve). Here we plot the converged results of simulations in the Fock space truncated at 7 quanta.}
\label{jump_probability}
\end{figure}

As soon as the irreversibly decoupled degree of freedom appeared, a quantum jump was performed.  Fig.\ref{jump_probability} presents the probability distribution of the quantum jumps $|c_q|^2$ (all possible choices). Here and below, the dependence of $c_q$ on all the earlier outcomes is not indicated.

The figure shows that for the integrable case at $K=0$, the probability distribution is very narrow, whereas for the chaotic regime $K=-10$ the jump probability distribution is very broad.

\begin{figure}[!h]
\centering
\includegraphics[width=0.5\textwidth]{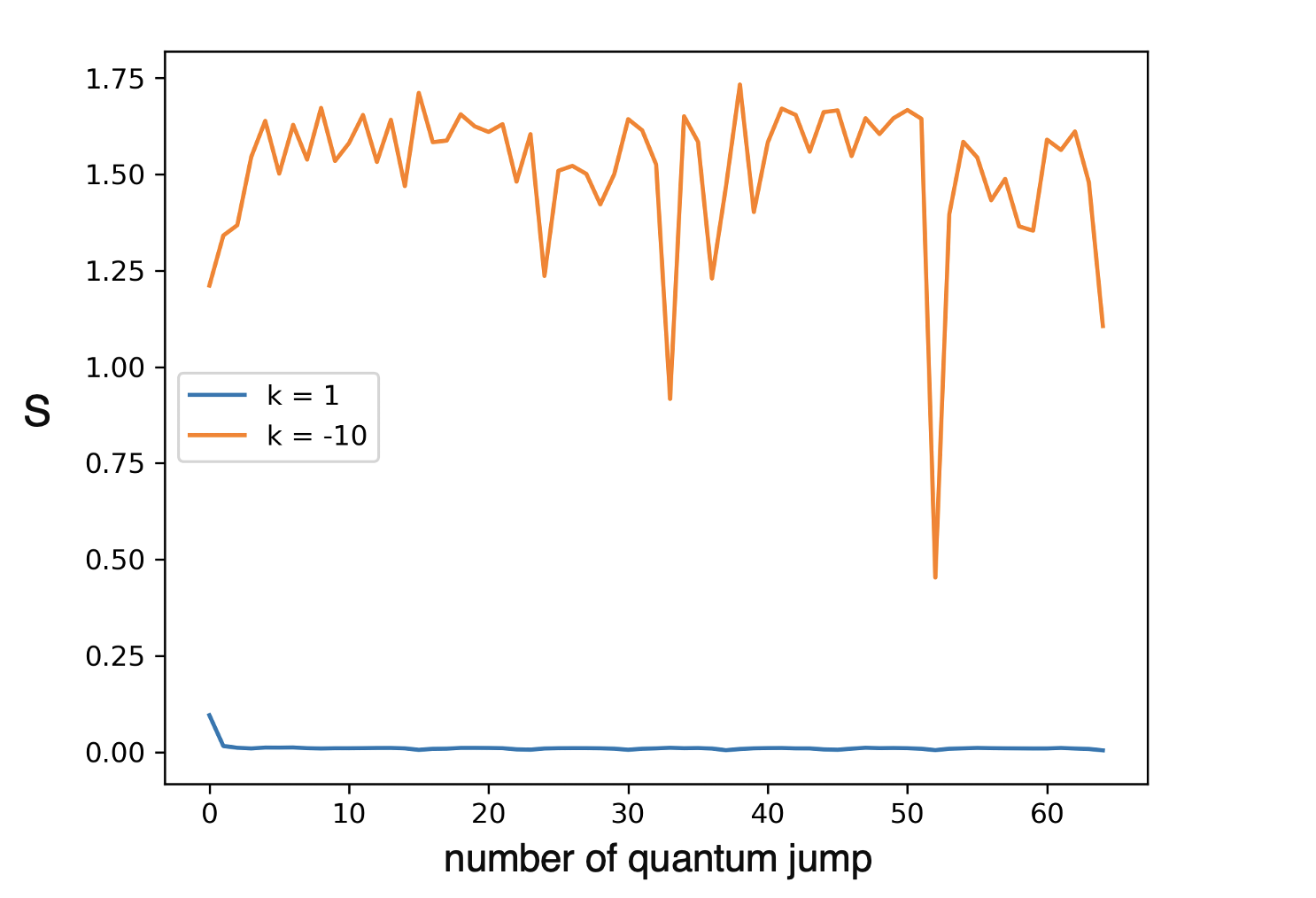}
\caption{Instantaneous entropy production along one trajectory for $K=1$ (blue curve) and $K=-10$ (orange curve).}
\label{prod_entropy}
\end{figure}

This procedure was repeated for the entire time interval $T$. One random implementation of the choice of quantum transitions was considered. In Fig.\ref{prod_entropy} the instantaneous production of entropy depends on the quantum jump number. In integrable and chaotic regimes, entropy along one trajectory behaves in radically different ways.

When $n$ quantum jumps have already happened and the moment of the next jump has come, we can expand the wave function of the system in terms of the Schmidt expansion (\ref{shmidt}) and from the previous set of significant modes, we select a new set of significant modes and a mode that is irreversibly decoupled (on which the projection is carried out):
\begin{multline}
|\Psi(t)\rangle = \sum_{q_{n+1}} c_{q_{n+1}} \times \\  \times 
|\Psi_{coll}^{(q_{n+1})}(t,q_1,...,q_n)\rangle_{rel} \otimes |\Psi_{J}^{(q_{n+1})}(t)\rangle_{\kappa_n^{out}}
\end{multline}

At the $n+1$ step, a new distribution of quantum jumps arises (a set of alternatives). The entropy for one jump increases:
\begin{equation}
    \Delta S = - \sum_{q_{n+1}} c_{q_{n+1}} \ln(c_{q_{n+1}})
\end{equation}

\begin{figure}[!h]
\centering
\includegraphics[width=0.5\textwidth]{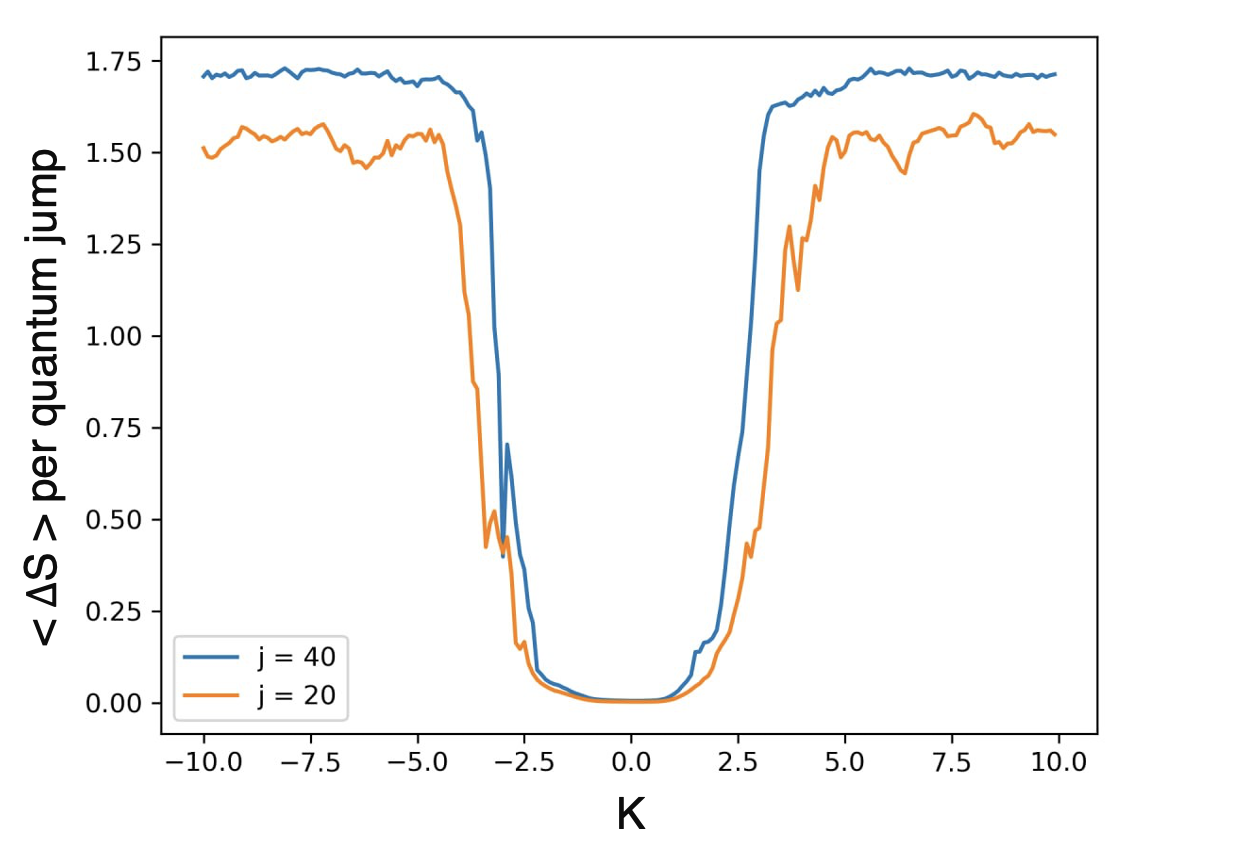}
\caption{Average entropy production (\ref{entropy_aver}) depending on kick strength K. One can see a sharp increase in entropy production in the region of crossover between integrable and chaotic dynamics. The calculation was performed for two different quantum numbers $j=20$ (orange curve) and $j=40$ (blue curve).}
\label{entropyFig}
\end{figure}

The average entropy production for one trajectory per quantum jump is:
\begin{equation}\label{entropy_aver}
\langle\Delta S\rangle = \dfrac{1}{N} \sum_N \Delta S
\end{equation}
where $N$ is the total number of quantum jumps. Fig.\ref{entropyFig} represents the average entropy production per quantum jump. As can be seen, its behaviour changes strongly when passing from the integrable case, where there is practically no increase in entropy, to the chaotic case, where there is a strong increase in entropy.

It was confirmed that in the integrable case the trajectories behave more regularly and the entropy practically does not increase, whereas in the transition to the chaotic case, the trajectories mix strongly and the entropy grows rather sharply. Moreover, with an increase in $j$, the entropy growth angle increases. Thus, it is assumed that the entropy production along one trajectory can be a criterion of quantum chaos.

\section{Conclusions\label{conclusion}}
The main idea was to introduce a definition of quantum chaos similar to the classical definition through the divergence of nearby trajectories.

Quantum trajectories can be introduced by connecting the system to the environment. In this case, the quantum environment is analogous to a recording device. The role of the information carrier in the quantum environment is played by the degrees of freedom that are irreversibly decoupled from the OQS (the stable records), which periodically arise during the evolution of the joint system in time.

In this work, we first offer a novel way of finding the degrees of freedom of the environment that carries information about the trajectory by averaging the OTOC. Second, on the basis of this, we introduce the definition of trajectories. As a criterion for quantum chaos, we propose using the entropy of the ensemble of given trajectories (\ref{entropy}).

Thus, one can consider the environment as a measuring device that autonomously selects the time of measurement and the preferred basis without the intervention of a human experimenter. In turn, during the evolution in real time, by measuring one after another irreversibly decoupled modes with a certain probability, a sequence of measurements is obtained, which results in a quantum trajectory.

It was confirmed that for regular motion, decoherent histories behave relatively regularly, whereas for chaotic motion, the recorded particle trajectory is more fluctuating and irregular. The entropy of an ensemble of such trajectories grows faster in the chaotic case than in the integrable case. It is also possible to observe a noticeable sharp increase in entropy during the transition of the system dynamics from integrable to chaotic at values of the kick strength $K$ from 2 to 3 (Fig.\ref{entropyFig}). Thus, this approach made it possible to fix the phenomenon of quantum chaos for the model of a quantum kicked top. We propose connecting any considered chaotic system to the environment and using the entropy of the ensemble of decoherent histories as a criterion of chaos.

We are currently developing a software package that implements the methods outlined in this work \cite{litlinkP}.

\subsection*{Acknowledgement}
This study was supported by the Roadmap for the Development of the High-Technology Field ``Quantum Computations'' no. 1/17654-D dated July 10, 2019 (State Atomic Energy Corporation Rosatom).

\end{document}